\documentclass[twocolumn]{aastex7}

\newcommand{\dif}{\mathrm{d}}
\usepackage{bm}

\usepackage{amsmath}
\usepackage{amssymb}
\usepackage{xcolor}
\usepackage{adjustbox}
\usepackage{CJK}
\usepackage{tablefootnote}

\shorttitle{Frequency of ZLK-driven Pollution}
\shortauthors{}
\graphicspath{{}{}}

\begin{document}
\begin{CJK}{UTF8}{gbsn}
\title{Constraining the Occurrence of ZLK-Induced White Dwarf Pollution with Dissipative Precession}

\author[0000-0002-4872-1021]{Isabella L. Trierweiler}
\affiliation{Department of Astronomy, Yale University, New Haven, CT 06511, USA}
% \correspondingauthor{Isabella L. Trierweiler}
\email{show}{isabella.trierweiler@yale.edu}

\author[0000-0002-4836-1310]{Konstantin Gerbig}
\altaffiliation{Eric and Wendy Schmidt AI in Science Fellow}
\affiliation{Department of Astronomy \& Astrophysics, University of Chicago, Chicago, IL 60637, USA}
\affiliation{Department of Astronomy, Yale University, New Haven, CT 06511, USA}
\email{kgerbig@uchicago.edu}

\author[0000-0002-7670-670X]{Malena Rice}
\affiliation{Department of Astronomy, Yale University, New Haven, CT 06511, USA}
\email{fakeemail2@google.com}

\begin{abstract}
Von Zeipel-Lidov-Kozai (ZLK) oscillations, induced by bound, perturbative companions to white dwarfs, have been suggested as a dynamical mechanism that may contribute to white dwarf pollution. 
To trigger ZLK oscillations, however, a 3-body system must reach a sufficiently large mutual inclination between orbits. The occurrence of these high-mutual-inclination configurations can be curtailed by dissipative precession at the protoplanetary disk stage, which pushes exoplanet-hosting close binary systems toward preferential orbit-orbit alignment.
In this work, we constrain the fraction of white dwarfs with binary companions that can undergo ZLK-driven pollution given the effects of dissipative precession. 
To accrete pollution via ZLK oscillations, a white dwarf binary system must be sufficiently inclined and the characteristic timescale of the oscillations must be sufficiently short to perturb material within the white dwarf's cooling age. 
Considering a sample of 4400 known white dwarf/main sequence binaries, we find that $50-70\%$ have favorable parameters for ZLK pollution, depending on the orbital separation of the polluting body. 
While the conditions for oscillations are favorable, the tendency for ZLK to result in massive but more infrequent polluters likely restricts the rates of ZLK-induced pollution among the observed population. 
In general, dissipative precession is a limiting factor in pollution rates for more closely separated binaries (initial separations $<500-800$~au), while ZLK timescale constraints are most limiting for wider binaries. 
\end{abstract}

%% Keywords should appear after the \end{abstract} command. 
%% The AAS Journals now uses Unified Astronomy Thesaurus (UAT) concepts:
%% https://astrothesaurus.org
%% You will be asked to selected these concepts during the submission process
%% but this old "keyword" functionality is maintained in case authors want
%% to include these concepts in their preprints.

\section{Introduction}\label{section:intro}
Polluted white dwarfs (WDs) offer valuable insights into the compositions of accreted planetary debris \citep[e.g.,][]{Jura2014, Xu2024}, enabling comparisons between the compositions of exoplanetary rocks/ices and material in the Solar System. 
However, the primary drivers of WD pollution remain under debate \citep{Veras2024}. 
Theorized mechanisms include planet-planet scattering \citep[e.g.,][]{Mustill2018, Maldonado2021, O'Connor2022, Veras2023}, radiation-driven migration \citep{Veras2022}, dynamical perturbations from post-main-sequence mass loss and planetary engulfment \citep{Smallwood2021, Akiba2024}, and perturbations from binary companions \citep{Bonsor2015, Stephan2017, Petrovich2017}.  

% pollution to WDs \citep{Mustill2018, O'Connor2022, O'Connor2023}. It is possible that such planets are more likely to be in high-multiplicity systems to induce instabilities more quickly \citep{Maldonado2021}. 

% On the extremely low-mass end, bodies as small as the Moon may also be capable of inducing pollution \citep{Veras2023}. As a final alternative, it is possible to induce pollution without any planet or stellar binary dynamics,  through mechanisms such as radiation-driven orbital decay \citep{Veras2022} or dynamical kicks from anisotropic mass loss during WD formation \citep{Akiba2024}. 

Binary-driven von Zeipel-Lidov-Kozai (ZLK) oscillations \citep{vonZeipel1910, Lidov1962, Kozai1962}, in particular, have been proposed as an efficient means of delivering especially massive and/or volatile-rich material to WDs, and maintaining accretion rates over Gyr timescales \citep{Stephan2017, Petrovich2017}. 
In this scenario, a distant companion star drives oscillations in the planets or debris belt bodies around the WD, exciting their eccentricities until they accrete onto the WD.
Quantifying the rate of binary-induced pollution versus other mechanisms can, therefore, aid in interpreting the source populations that are sampled by WD pollution, with important implications for understanding how the materials comprising extrasolar systems compare with the solar system. 

A baseline requirement for a 3-body system to undergo ZLK oscillations is that it must begin with a sufficiently large mutual inclination ($>39.2^\circ$) between orbits \citep{Naoz2016}. 
In the case of pollution on a WD with a stellar binary companion, this mutual inclination is measured between the orbital plane of the body that will pollute the WD -- for example, a planetary companion or debris disk -- and the orbital plane of the wide-orbiting companion star. 
Among main-sequence (MS) exoplanet-hosting wide binary systems, several studies \citep{Christian2022,Dupuy2022, Lester2023, Rice2024, Christian2025} have shown that the degree of observed orbital alignment between the planetary orbital plane and the companion star is strongly dependent on the separation between the planet host and companion star.  
These studies find that stellar binary systems separated by $\lesssim 800$ au demonstrate a trend towards alignment, whereas more widely separated binaries ($\gtrsim 800$ au) are consistent with random orientations. 

The population-level tendency toward alignment for close- to moderate-separation binaries may be attributed to energy dissipation in the systems' natal protoplanetary disks as they precess under the perturbing gravitational influence of a stellar companion \citep[][]{Lai2014, foucart2014evolution, Zanazzi2018}.
For sufficiently close binaries, this dissipative precession drives primordially misaligned systems toward alignment \citep{Gerbig2024}, producing a population of systems that would be less prone to ZLK dynamics than otherwise expected.

The goal of this work is to derive an upper limit for the rate at which WDs in WD/MS binaries can be polluted by ZLK oscillations with a binary star companion -- incorporating, for the first time, the influence of dissipative precession in shaping the system's early geometry.
To accomplish this, we calculate the fraction of known WD/MS binaries with (1) sufficiently large orbit-orbit misalignments at the end of disk dispersal (mutual inclination $>39.2^\circ$) and (2) ZLK timescales shorter than the typical cooling ages of observed WDs. We extrapolate the initial separations between the WD progenitor and MS companion and apply the model of \cite{Gerbig2024} to evolve the system dynamics during the protoplanetary disk phase. 

This paper is organized as follows. We outline our sample selection criteria in Section \ref{methods:sample} and calculations of alignment generation via dissipative precession in Section \ref{section:methods}. We present our results in Section \ref{section:results} and discuss the implications for pollution candidates in Section \ref{section:discussion}. Finally, we share our conclusions in Section \ref{section:conclusion}. 

\section{Sample of WD/MS binaries}\label{methods:sample}

%We draw our sample from the \textit{Gaia} DR3 catalog \citep{GaiaCollaboration2023}, which provides projected separations for over $10^4$ WD/MS binary systems. 

\subsection{Initial sample selection}
We select a sample of WD/MS binaries by cross-referencing the binary systems identified from \textit{Gaia} EDR3 astrometry \citep{GaiaCollaboration2023} in \citet{El-Badry2021} with sources identified as white dwarfs in \citet{GentileFusillo2021}. We consider only pairs where the RUWE parameter (astrometric goodness of fit) for both binary components is $<1.4$ and the $R$ factor describing chance alignment in \citet{El-Badry2021} is $<0.1$. 

We also require the WD component to have a mass modeled by \cite{GentileFusillo2021}; these are only reported  by \cite{GentileFusillo2021} if the modeled mass falls in the WD mass range of 0.1-1.4$~\mathrm{M_\odot}$. Three separate atmosphere models are applied to each WD to calculate stellar parameters: pure H, pure He, and a mix of H of He (with the ratio H/He  = $10^{-5}$). These masses are shown in Figure \ref{fig:mass_sep_dist}. The results of each model are only reported if the corresponding stellar parameters (temperature and $\log g$) fall in valid ranges determined by \cite{GentileFusillo2021}. As a result, while every WD in our sample has a mass for the pure H model, not every star has a corresponding mass reported for the pure He and mixed models. Masses determined by the three methods can differ by up to about $1~\mathrm{M_\odot}$, with a median discrepancy of 0.5~$\mathrm{M_\odot}$. We carry out separate calculations for each case throughout this work. 

\subsection{Sub-samples of white dwarfs}
Within this larger sample of WD/MS binaries, we also consider two sub-samples. Both of these sub-samples are relatively small in number, so we use them only to benchmark our model for the full sample of 4400 WDs (described in Section \ref{methods:initialconditiona}).

The first sub-sample is the volume-limited 40~pc survey of WDs as reported in \cite{O'Brien2024}. The 40 pc survey is estimated to be $>99\%$ complete, serving as a benchmark against which to compare the masses and stellar properties of our larger sample of binary WDs. 
25 of the WDs in our binary sample are within the 40~pc sub-sample. \cite{O'Brien2024} provides ``corrected" masses for these WDs, where an opacity correction is applied to the mass modeling to address discrepancies in mass fitting for low-mass, cool ($<6000$ K) WDs. The median ``corrected" mass of the 40~pc WDs is $\approx0.6~\rm M_\odot$. The ``corrected" masses are most similar to the pure H masses from  \cite{GentileFusillo2021}, with a median difference of $0.02~\mathrm{M_\odot}$. Binary parameters---including initial stellar mass ratios and semi-major axes---obtained using the ``corrected" masses are calculated following the same methods as the H, He, and mixed masses from \cite{GentileFusillo2021} (methods described in Section \ref{section:methods}). 

The second sub-sample is the subset of binary systems with a WD component which is polluted, selected by cross-referencing the binary sample with the sample of polluted WDs compiled by \cite{Williams2024}. For this work, we consider any WD with a Ca detection to be polluted. Ca is one of the most frequently observed signs of WD pollution, and we therefore select this tracer to maximize our sample of polluted WDs across a variety of binary separations. Nonetheless, only 11 of the WDs in our sample have measured Ca abundances. 11/4400 is almost certainly an underestimate of the overall pollution rate for WDs in moderate-separation binaries, driven by observational limits and incomplete coverage. In the more complete 40~pc survey, \cite{O'Brien2024} find 2/10 binary WDs are polluted (though both of these are interior to our considered range of semi-major axes).

\section{Methods}\label{section:methods}

In this section we describe the semi-analytic models deployed in this work. In Section \ref{methods:initialconditiona} we calculate the present-day orbital parameters of the binaries, and we use these to extrapolate the initial (prior to WD formation, while both stars are still on the MS) orbital parameters. 
In Section \ref{section:simulation_methods}, we describe our method to calculate the evolution of these initial orbital parameters under dissipative precession during the protoplanetary disk phase using the model of \cite{Gerbig2024}. 
Finally, we describe our adopted ZLK timescale constraints in Section \ref{section:kozai_time}. 

\subsection{Initial conditions of binary orbits}\label{methods:initialconditiona}

We calculate the present-day semi-major axis of each WD/MS binary as \citep{Torres1999}
\begin{equation}
    a_{\rm b, WD} = \frac{s}{1 - e_{\rm b}\cos(E)} \frac{1}{\sqrt{1 - \sin^2(\omega+\nu_T) \sin^2 i}},
\end{equation}
\noindent where $s$ is the projected separation, $e_{\rm b}$ is the eccentricity, $E$ is the eccentric anomaly, $\omega$ is the longitude of periastron, $\nu_T$ is the true anomaly, and $i$ is the projected inclination (note that $i$ is distinct from the true mutual inclination, which we use throughout this work to measure the alignment of the binary system). 

An orbital eccentricity value is needed to translate each \textit{Gaia}-derived projected binary separation into an inferred present-day orbital semi-major axis. As a simple approximation, we assume that eccentricities are uniformly distributed between zero and one and that the eccentricity of the binary is maintained through the formation of the WD (see Section \ref{section:binary_eccentricity} for a discussion on the implications of this assumption). While a uniform eccentricity distribution is typically accurate for close-separation stellar binaries (separations $\lesssim100$~au), the distribution may steepen to include a higher fraction of eccentric pairs at wide separations \citep{Hwang2022, Wu2025}. We discuss the implications of more complex eccentricity distributions on ZLK-induced pollution rates in Section \ref{section:binary_eccentricity}.  The projected inclination $i$ and all other orbital angles are randomly selected assuming a uniform distribution between 0 and $2\pi$. 

Assuming conservation of angular momentum, the semi-major axis of the binaries during the main sequence is then
\begin{equation}\label{equation:aMS}
    a_\mathrm{b, prog} = a_{\rm b, WD} \left( \frac{M_{\rm WD}}{M_{\rm prog}}\right)^2 \frac{M_{\rm C} + M_{\rm prog}}{M_{\rm C} + M_{\rm WD}},
\end{equation}
\noindent where $a_{\rm b, WD}$ is the present-day semi-major axis calculated from separations reported by \cite{El-Badry2021}, $M_{\rm WD}$ is the mass of the WD, $M_{\rm prog}$ is the WD's progenitor mass, and $M_{\rm C}$ is the mass of the companion.

We characterize the masses of the MS companions ($M_{\rm C}$) using the \textsc{isochrones} package \citep{Morton2015} to fit the \textit{Gaia} photometry and parallax of each companion. To calculate initial masses of the WD progenitors under each of the atmosphere assumptions, we use the empirical initial-final mass relation from \cite{Cunningham2024}, which was derived from the 40~pc survey of WDs.
This relation is valid for WD masses $>0.5~\mathrm{M_\odot}$; we therefore do not include any initial masses calculated from current WD masses $>0.5~\mathrm{M_\odot}$ in our sample. Within the initial-final mass relation, the uncertainty in a calculated progenitor mass is typically $\approx0.25~\mathrm{M_\odot}$ for a given WD mass. 

In this work, we focus on the sample of binaries with calculated initial separations ($a_\mathrm{b, prog}$) within 2500~au. Beyond this limit, the characteristic timescale of ZLK oscillations for an average system in our sample (WD mass of 0.6~$M_\odot$, companion of 0.4~$M_\odot$, and a planet at 10~au) becomes prohibitively long ($\gg1$~Gyr) for eccentricities less than 0.5. Other mechanisms, such as galactic tides, may instead become the more relevant binary-dependent drivers of pollution at these wide separations \citep{Bonsor2015}. We also remove binary pairs with initial separations within 200~au, as planet formation may be repressed in these close binaries \citep{Moe2021} and the associated protoplanetary disk timescales in our model become very short (see Eq.~\ref{equation:rdisk}). Our final sample of moderately-close binaries consists of $\approx4400$ pairs. 

\begin{figure*}
        \centering
        \includegraphics[width=\linewidth]{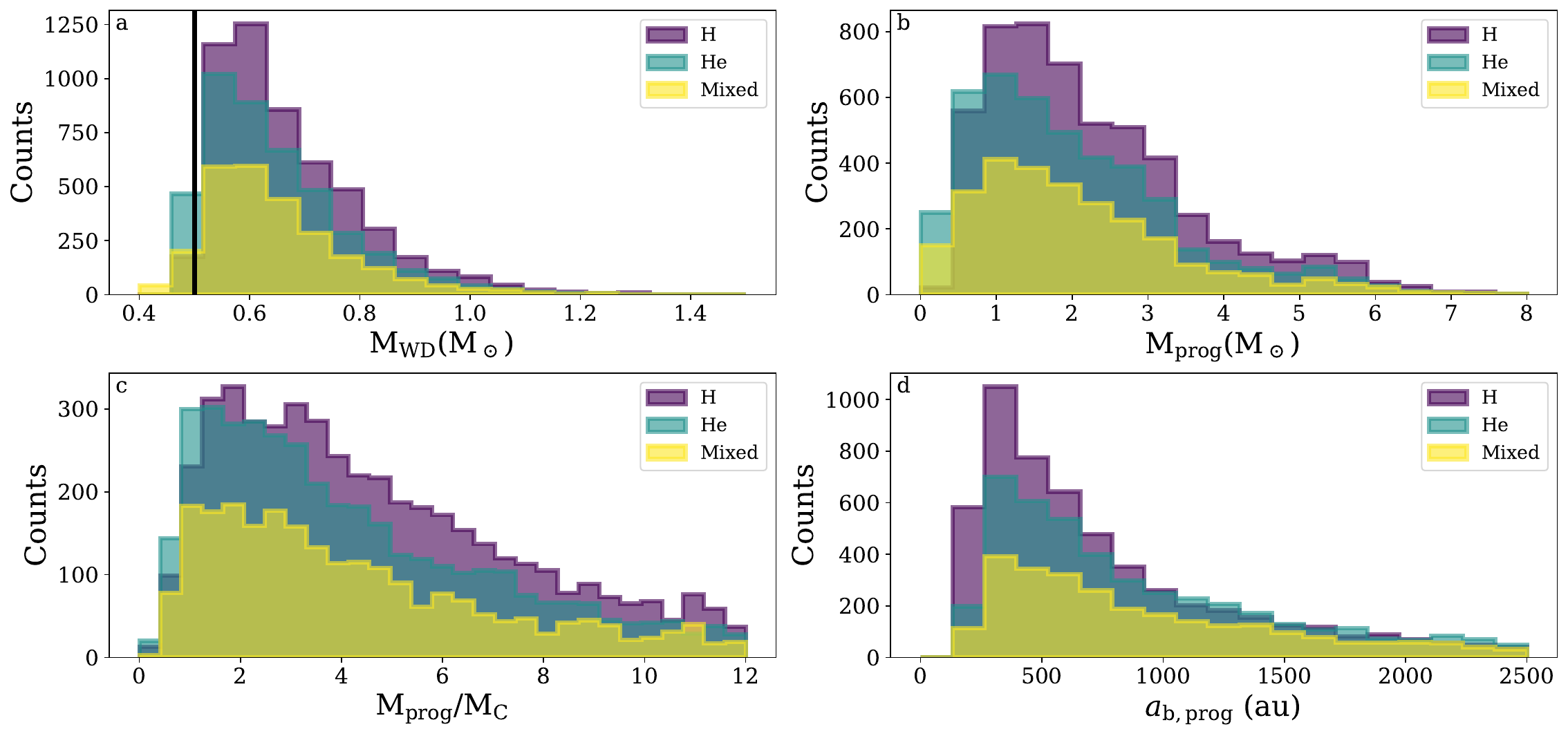}
        \caption{Panel a: Masses for the WD component of each binary pair, under the assumptions of pure H, pure He, and mixed H/He atmospheres \citep{GentileFusillo2021}. Initial-final mass models typically do not extend lower than $0.5 \mathrm{M_\odot}$ (the vertical line); we therefore include only WDs where at least one of the atmospheric solutions includes a mass $\geq 0.5 \mathrm{M_\odot}$. \\
        Panel b: Distribution of initial (progenitor) masses for the WD components of the binaries, obtained by applying the initial-final mass relation from \cite{Cunningham2024} to the WD masses in Panel a.  \\ 
        Panel c: Distribution of the ratios of initial masses for the WD components relative to the masses of their MS companions. This corresponds to the mass ratio of the binary pairs when both components were on the main sequence (before formation of the WD). The companion masses are obtained through isochrone fitting (see Section \ref{methods:initialconditiona}).\\
        Panel d: Distribution of the initial semi-major axes of the WD/MS binary pairs when both components were on the main sequence. The choice of pure H, pure He, or mixed H/He mass model affects both the progenitor mass of the WD and the initial semi-major axis via Eq.~\ref{equation:aMS}. \\
        The sample of WDs in the mixed H/He scenario is much smaller than those of the pure H and He cases, as \citealt{GentileFusillo2021} reports mixed H/He results for a narrower range of effective temperatures ($\gtrsim6500$~K) compared to the pure H and He cases ($\gtrsim4000$~K).}
        \label{fig:mass_sep_dist}
\end{figure*}

Figure \ref{fig:mass_sep_dist} shows the calculated mass ratio and semi-major axis for each binary pair prior to WD formation. We use these initial conditions to evolve each binary pair through disk dissipation following the approach detailed in Section \ref{section:simulation_methods}. 

\subsection{Dissipative precession calculations}\label{section:simulation_methods}

To assess the probability that a given WD/MS system is aligned and thus cannot undergo ZLK oscillations, we use the initial masses and orbital parameters obtained in the previous section to conduct a suite of calculations of the dynamical evolution during the WD progenitors' protoplanetary-disk-hosting phases. 
We draw the masses for the WD progenitor ($M_\mathrm{prog}$) and binary companion ($M_\mathrm{C}$), the initial semi-major axis ($a_\mathrm{b, prog}$), and the eccentricity ($e_\mathrm{b}$) from the sample of WD/MS binaries, using the median of the values obtained from the three WD atmospheric models. 
The physical parameters required as inputs for the dissipative precession model are derived from $(M_\mathrm{prog}, M_\mathrm{C}, a_\mathrm{b,prog}, e_\mathrm{b})$ using the scaling relations described below. 
For each WD progenitor system, we calculate the dynamical evolution for a range of initial mutual inclinations $\theta_{\rm db}(t=0)$ selected from 0 to $\pi/2$ radians, to obtain a distribution of final mutual inclinations for the selected set of binary parameters $(M_\mathrm{prog}, M_\mathrm{C}, a_\mathrm{b,prog}, e_\mathrm{b})$. 
% Finally, we sum all calculations and bin by initial semi-major axes $a_\mathrm{b,prog}$, which allows us constrain the expected alignment fraction of WD progenitor systems as a function of binary separation. 

We model the secular evolution of the progenitor system by solving the coupled dynamical equations for the progenitor disk angular momentum vector and the progenitor's spin vector following \citet{Zanazzi2018}. Specifically, we consider an effectively flat disk that rigidly precesses about the assumed-constant binary orbital normal with frequency $\omega_\mathrm{db}$ -- see Appendix~\ref{section:appenidx_dissipative_precession} for details. The disk drives precession of the stellar spin ($\omega_\mathrm{sd}$), and conversely, the star drives precession in the disk $\omega_\mathrm{ds}$. The disk's mass loss causes $\omega_\mathrm{sd}$, initially the fastest rate, to fall below the constant $\omega_\mathrm{db}$. This resonant crossing can tilt, or even flip, the stellar spin vector, exciting the mutual disk - stellar spin inclination \citep[e.g.,][]{Lai2014}. In addition, and crucially, the dynamical equations include damping terms that are associated with internal torques induced by small warps in the disk profile \citep{Zanazzi2018}. This viscosity-dependent dissipation is necessary for the progenitor system to move towards alignment. 

The damping rate of the progenitor system via dissipative precession depends quite sensitively on system properties, for which we employ the scaling relations in \citet{Gerbig2024} as well as additional ones described below to further reduce the number of free parameters. Following \citet{Gerbig2024}, we choose an initial mass of the disk $M_\mathrm{d,0} = 0.3 M_{\rm prog}$, and it is assumed that the disk undergoes homologous mass loss described by 
\begin{align}
    M_\mathrm{d}(t) = \frac{M_\mathrm{d,0}}{1+t/t_\mathrm{v}},
\end{align}
where the viscous timescale $t_\mathrm{v}$ is taken to scale as \citep[Eq. (27) in][]{Gerbig2024}
\begin{align}
    t_\mathrm{v} = 0.5 \mathrm{\ Myr} \cdot \left(\frac{\alpha}{0.05}\right)^{-1} \left(\frac{M_{\rm prog}}{M_\odot}\right)^{\frac{3}{14}}\left(\frac{r_\mathrm{out}}{100~\mathrm{au}}\right)^{\frac{19}{20}}.
\end{align}
Here, we introduced the disk's outer truncation radius $r_\mathrm{out}$ and the Shakura-Sunyaev  $\alpha$-parameter \citep{Shakura1973} as a stand-in for the disk's viscosity. 

Acceptable values for $\alpha$ can span orders of magnitude, and  viscosity appears observationally uncorrelated with global properties \citep{Rafikov2017}. We thus employ a fiducial viscosity scaling relation in which a floor motivated by turbulence due to the Magnetorotational Instability (MRI) with $\alpha_\mathrm{MRI} = 10^{-3}$ \citep{Flock2017} transitions to a higher, Gravitational Instability (GI) - driven contribution of $\alpha_\mathrm{GI} = 0.06$ \citep{Rafikov2015} at larger masses:
\begin{align}
    \alpha = \alpha_\mathrm{MRI} + \frac{\alpha_\mathrm{GI}-\alpha_\mathrm{MRI}}{2}\left[1+\tanh\left(\frac{M_{\rm prog} - M_\mathrm{break}}{M_\mathrm{width}}\right)\right].
\end{align}
We choose $M_\mathrm{break} = 1.5M_\odot$ and $M_\mathrm{width} = 0.5 M_\odot$ to characterize the location and width of the transition regime, respectively. For the disk's outer edge, we assume a simple scaling with disk mass and the radius of the WD progenitor $R_{\rm prog}$ (Eq.~\ref{equation:r_prog}), motivated by empirical sub-linear size-luminosity relations from ALMA continuum surveys \citep[e.g.][]{Hendler2020}, i.e.,
\begin{align}
    r_\mathrm{out,base} = 120 \mathrm{\ au} \cdot \left(\frac{M_\mathrm{d,0}/M_\odot}{0.01}\right)^\frac{1}{4}\left(\frac{R_{\rm prog}}{R_\odot}\right)^{\frac{1}{2}}.
\end{align}
We further truncate the disk at \citep{Miranda2015}
\begin{align}\label{equation:rdisk}
    r_\mathrm{tide} \approx 0.35 \left(1-e_\mathrm{b}\right)a_\mathrm{b, prog}
\end{align}
to account for tidal stripping such that
\begin{align}
    r_\mathrm{out} = \mathrm{min}\left(r_\mathrm{out,base}, r_\mathrm{tide}\right).
\end{align}
Assuming an equilibrium disk with a temperature scaling of \citep{Chiang1997, Wu2021}
\begin{align}
    T(r) = 140\mathrm{\ K}\cdot \left(\frac{M_{\rm prog}}{M_\odot}\right)^{-\frac{1}{7}}\left(\frac{L_{\rm prog}}{L_\odot}\right)^{\frac{2}{7}} \left(\frac{r}{1~\mathrm{au}}\right)^{-\frac{9}{20}},
\end{align}
and a mass-luminosity relation of $L_{\rm prog} \propto M_{\rm prog}^{3/2}$ \citep[see e.g.,][]{Chachan2023}, the disk aspect ratio at the outer disk edge scales as \citep{Gerbig2024}
\begin{align}
    h_\mathrm{out} = 0.08 \cdot \left(\frac{M_{\rm prog}}{M_\odot}\right)^{-\frac{5}{14}} \left(\frac{r_\mathrm{out}}{100~\mathrm{au}}\right)^{\frac{11}{40}}.
\end{align}
We approximate the WD progenitor's radius $R_{\rm prog}$ using a main-sequence mass-radius relation $R_{\rm prog}\propto M_{\rm prog}^{0.8}$ \citep[e.g.,][]{Feiden2012} with an additional boost factor to account for pre-main-sequence inflation, where stars are typically 2-5 times larger than their zero-age main sequence counterparts \citep[][]{Stahler1988}, i.e.
\begin{align}\label{equation:r_prog}
    R_{\rm prog} = R_\odot \left(\frac{M_{\rm prog}}{M_\odot}\right)^{0.8} \left(2+\frac{3}{1+(M_{\rm prog}/M_\odot)^2}\right)
\end{align}
For the disk's inner radius, we simply take $r_\mathrm{in} = 7~R_{\rm prog}$, which is consistent with magnetospheric truncation models in young stellar objects \citep[see e.g.,][]{Tofflemire2017}. We invoke photoevaporative mass loss as the cause of the disk's ultimate dispersal and set the disk's lifetime equal to the photoevaporation time \citep[e.g.,][]{Komaki2021}
\begin{align}
    t_\mathrm{photo-evap} = 10 \mathrm{\ Myr} \cdot \left(\frac{M_{\rm prog}}{M_\odot}\right)^{-1}
\end{align}
while further enforcing
\begin{align}
    t_\mathrm{life} = \max\left[0.3 \mathrm{\ Myr}, \min\left(t_\mathrm{photo-evap}, 10\mathrm{\ Myr}\right)\right]
\end{align}
to keep the integration time within a reasonable range. %In addition, we fix the following parameters to the fiducial values in \citet{Gerbig2024} for all of our models: the stellar rotation rate; the stellar quadrupole coefficient and spin normalization coefficient; and the power-law index $p = 1$ of the disk's surface density profile, i.e. $\Sigma(r) \propto r^{-p}$. 

This leaves four free parameters which we randomly draw, without replacement, from the calculated masses and orbital parameters of the observed binaries: the white dwarf progenitor (primary) mass $M_{\rm prog}$, the binary companion mass $M_\mathrm{\rm C}$, the binary separation $a_\mathrm{b, prog}$, and the binary eccentricity $e_{\rm b}$. The binary eccentricity, which was set to $e_\mathrm{b}=0$ in \citet{Gerbig2024} and is assumed to follow a uniform distribution in this work, primarily acts to decrease the effective binary separation in the small warp limit, i.e. $a_\mathrm{b, eff} = a_\mathrm{b, prog}\sqrt{1-e_{\rm b}^2}$, thus increasing the disk's precession rate and the damping rate. 

\subsection{ZLK timescale and WD cooling age comparison}\label{section:kozai_time}

For a given WD/MS system that can undergo ZLK oscillations, the WD can only experience ZLK-driven pollution in its lifetime if the characteristic timescale for ZLK oscillations is shorter than the cooling age of the WD. 

We calculate cooling ages ($T_{\rm cool}$) for the WDs in our sample by interpolating the model grids by \cite{Bedard2020} (as integrated into the Montreal White Dwarf Database, \citealt{Dufour2017}) based on the log gravity and effective temperature values provided by \cite{GentileFusillo2021} for each atmosphere model. The median cooling ages for the three atmosphere types (H, He, mixed) are 2.52, 2.77, and 1.26 Gyr, respectively. The relatively lower ages for the `mixed' WDs are primarily because the mixed model is only applied to WDs with effective temperatures above about 6500~K. 

We calculate the ZLK timescale for a body around a WD with a distant MS companion as \citep{Kiseleva1998} 
\begin{equation}\label{eq:kozaitime}
    T_{\rm ZLK} = \frac{\sqrt{M_{\rm WD}}}{M_{\rm C}} \frac{a_\mathrm{b, WD}^3}{a_{\rm p}^{3/2}} (1-e_{\rm b}^2)^{3/2},
\end{equation}
\noindent where $M_{\rm WD}$ is the WD mass, $M_{\rm C}$ is the mass of the companion, $a_\mathrm{b, WD}$ and $e_{\rm b}$ are the semi-major axis and eccentricity of the stellar binary, and $a_{\rm p}$ is the semi-major axis of the body orbiting the WD. 

$T_{\rm ZLK}$ depends strongly on the choice of $a_{\rm p}$. 
Closer companions would drive $T_{\rm ZLK}$ up, reducing the number of WDs that could be polluted within their cooling ages, while more distant companions increase the number of pollutable WDs. 
For material to survive post-main-sequence evolution, the smallest orbit for a polluting body in the WD phase is approximately $a_{\rm p}=$10~au. 
However, if the polluting bodies originate on wide orbits in the outer system, such as from a Kuiper Belt analog, $a_{\rm p}$ in the WD phase could be up to hundreds of au. 
To test the impact of this variation on modeled pollution rates, we consider $a_{\rm p}=10$~au and $a_{\rm p}=100$~au as the two semi-major axis limits of polluting bodies. We choose these limits as rough approximations for the post-MS, widened orbits of inner- and outer-solar-system analogues.

\section{Results}\label{section:results}   

\subsection{Pollution criteria}\label{section:pollution_criteria}

The fraction of binary systems that can undergo pollution via ZLK oscillations depends on two criteria. First, the system must be sufficiently misaligned, with mutual inclination $>39.2^\circ$, to trigger ZLK oscillations. 
Second, the characteristic timescale for ZLK oscillations (Section \ref{section:kozai_time}) must be sufficiently short to perturb material within the calculated cooling age of the WD ($T_{\rm ZLK} < T_{\rm cool}$). 
We apply these constraints to the results of our simulation in following sections.

\subsection{Modeled alignment fractions}\label{section:simulation_results}

For each WD/MS binary, the dissipative precession calculations described in Section \ref{section:simulation_methods} result in a distribution of mutual inclinations after the dispersal of the protoplanetary disk. We assume that the resulting mutual inclination remains constant through the formation of the WD. To derive a general function describing the probability of alignment for a binary of a given initial semi-major axis ($f_{\rm aligned}(a_\mathrm{b, prog})$), we sum across the distributions for individual WD/MS systems and calculate the total fraction of final mutual inclinations smaller than the critical ZLK angle of $39.2^\circ$. 
% In this section we outline the dependencies of $f_{\rm aligned}(a_\mathrm{b, prog})$. 

\begin{figure}
    \centering
    \includegraphics[width=1\linewidth]{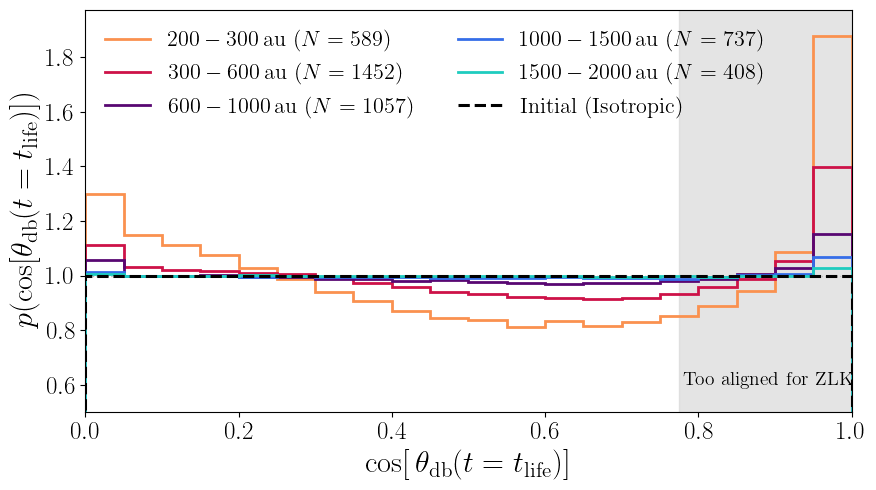}
    \caption{Distributions of final mutual inclinations $\theta_{\rm db}(t = t_\mathrm{life})$ for the simulated systems at the end of disk dissipation. Results are binned by color according to the initial separation between the WD progenitor and MS companion. The gray shaded region indicates where ZLK oscillations are prevented by mutual inclinations of $\theta_\mathrm{db}(t = t_\mathrm{life}) <= 39.2^\circ$ corresponding to $\cos[ \theta_\mathrm{db}(t = t_\mathrm{life})] \geq 0.775$.}
    \label{fig:sim_alignment}
\end{figure}

In Figure \ref{fig:sim_alignment}, we bin the binary sample by initial semi-major axis ($a_{\rm b, prog}$) and show the summed distribution of each bin's final mutual inclinations, visualized as the distribution of $\cos[ \theta_{\rm db}(t = t_\mathrm{life})]$. 
For large separations $a_{\rm b, prog}\gtrsim800$~au, the presence of the stellar companion has a minimal effect on the degree of alignment, so all values for $\cos \theta_{\rm db}(t = t_\mathrm{life})$ are equally likely according to the initial isotropic distribution. 
At closer separations, the stellar companion can preferentially align systems such that up to a third of the population with $a_{\rm b, prog}\lesssim800$~au can reach alignments of $ \theta_{\rm db}(t = t_\mathrm{life}) < 39.2^\circ$, or $\cos [\theta_{\rm db}(t = t_\mathrm{life})] \geq \cos (39.2^\circ) = 0.775$, by the time the disk dissipates. 

\begin{figure}
    \centering
    \includegraphics[width=\linewidth]{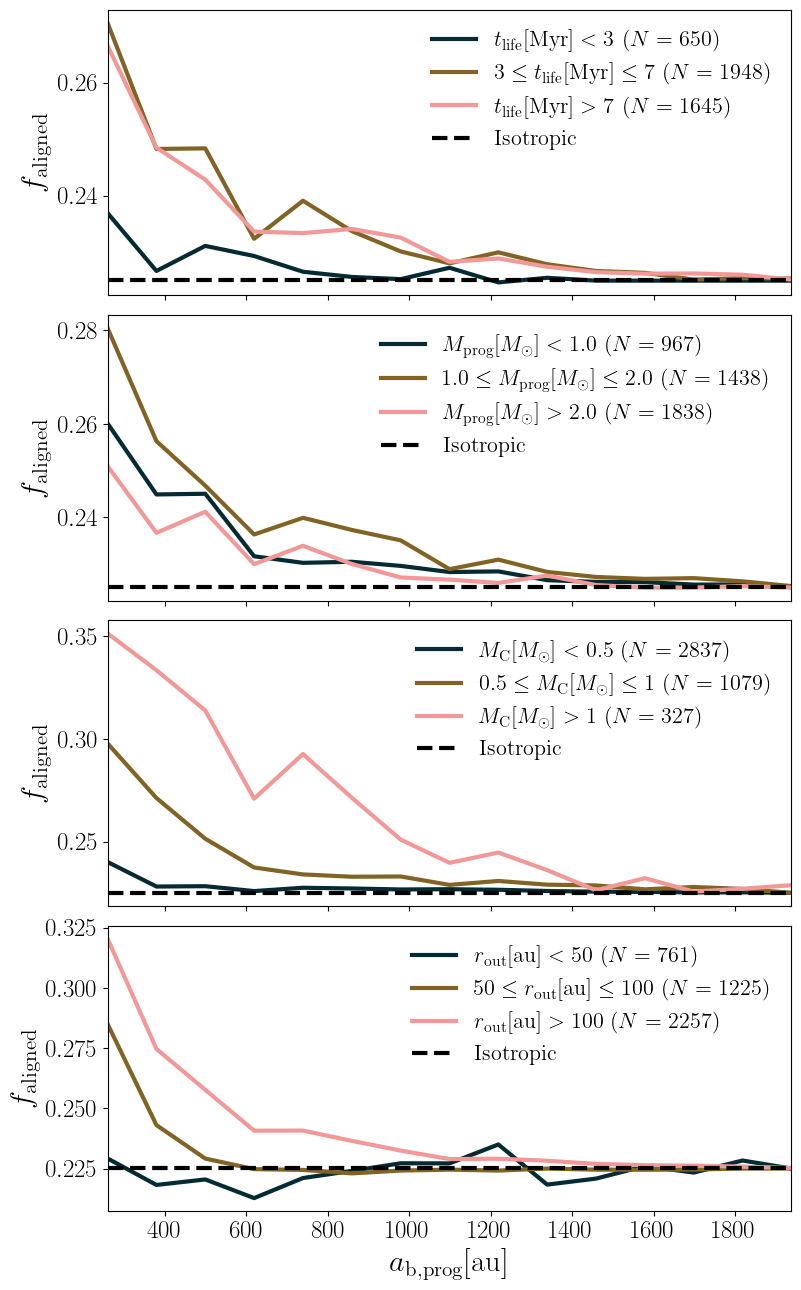}
    \caption{Fraction of systems that will not undergo ZLK oscillations due to alignment generated by binary-driven dissipative precession $f_\mathrm{aligned} \equiv f[\theta_\mathrm{db}(t=t_\mathrm{life}) < 39.2^\circ]$, as a function of initial binary separation $a_\mathrm{b, prog}$ obtained via Eq.~\eqref{equation:aMS}. From top to bottom, the four panels separate the samples according to the disk lifetime, WD progenitor mass, MS companion mass, and disk outer truncation radius. }
    \label{fig:sim_align_summed}
\end{figure}

Taking the distribution of final mutual inclinations for simulated binaries, we then calculate the fraction of systems that conclude their protoplanetary disk phases with a mutual inclination smaller than the critical ZLK angle of $39.2^\circ$. 
Figure \ref{fig:sim_align_summed} shows this fraction of binaries which cannot undergo ZLK oscillations versus the initial semi-major axis $a_{\rm b, prog}$, dividing systems by disk lifetime, WD progenitor mass, MS secondary mass, and disk outer truncation radius. 
We show results only out to 2000~au, as the alignment fraction quickly falls to the isotropic baseline with increasing initial semi-major axis. 

We find that companion mass has the strongest effect on the fraction of aligned systems in a given bin of $a_{\rm b, prog}$, as more massive companions drive more efficient inclination damping \citep{Gerbig2024}. 
Variations in WD progenitor mass and disk lifetime have a less pronounced impact on the fraction of aligned systems. 
Finally, dissipative precession is most efficient for more extended disks, so that binaries with disks of small radii ($<50$~au) tend to follow the isotropic baseline. 

As a whole, the modeled systems show much stronger alignment at close initial semi-major axes ($a_{\rm b, prog}\lesssim 800$ au), consistent with both the results of \citet{Gerbig2024} and the observed trends in alignment for exoplanet-hosting binaries \citep[e.g.,][]{Christian2022}. At the closest end ($200-300$ au), up to a third of systems end up too aligned to allow for ZLK oscillations. On the wider end, beyond $a_\mathrm{b, prog} \approx 1000$~au, alignment fractions relax toward the isotropic baseline of $\approx 22\%$. The effects of dissipative-precession-induced alignment weaken with decreasing disk lifetime $t_\mathrm{life}$, increasing primary mass $M_{\rm prog}$, decreasing secondary mass $M_{\rm C}$, and decreasing disk radius $r_\mathrm{out}$.

Because the efficiency of dissipative precession varies most strongly with companion mass, we use the third panel of Figure \ref{fig:sim_align_summed} to fit a general alignment probability function $f_{\rm aligned}(a_\mathrm{b, prog})$. We group the binary sample by companion mass, bin by initial semi-major axis in increments of 50~au, and use the  \texttt{curve\_fit} function from \texttt{SciPy} to fit an exponential to each curve. Our resulting alignment probability function is 
\begin{equation}\label{eq:prob_func}
    f_{\rm aligned}(a_\mathrm{b, prog}) = Ae^{-Bx}+0.225,
\end{equation}
where

\begin{align} 
&M_{\rm C}/M_\odot<0.5: \nonumber\\  
&\quad \quad A = 0.1758\pm0.0969, B=0.0118\pm 0.0026\nonumber\\
&0.5<M_{\rm C}/M_\odot<1.0:\nonumber\\  
& \quad \quad A = 0.1907\pm 0.0138, B=0.0045\pm 0.0003\nonumber\\
&M_{\rm C}/M_\odot>1.0: \nonumber\\
&\quad \quad A = 0.2164\pm 0.0154, B=0.0022\pm 0.0002\nonumber.
\end{align}

\subsubsection{Alignment of the binary sample}\label{section:alignment_full}

We now apply the alignment probability function (Eq.~ \ref{eq:prob_func}) to the different atmospheric mass models for our full sample (H, He, and mixed), as well as the 40~pc sub-sample, to calculate the fraction of these systems that should be aligned at an early stage due to dissipative precession. 

\begin{figure}
    \centering
    \includegraphics[width=1\linewidth]{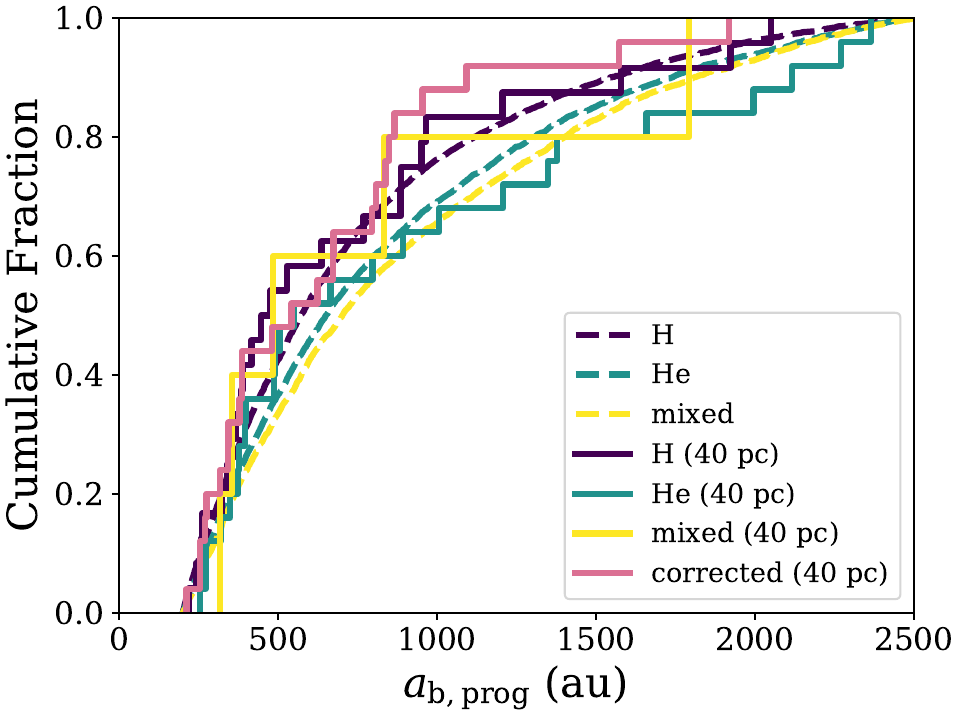}
    \caption{Cumulative distribution of initial semi-major axes, determined from WD progenitor masses, for the 40~pc sub-sample of binaries (solid lines), compared to analogous results for our full sample drawn from \textit{Gaia} (dashed lines). We show results for pure H, He, and mixed H/He atmosphere cases for both the full sample and 40~pc sub-sample, as well as results using the ``corrected" masses from \cite{O'Brien2024} for the 40~pc sub-sample.}
    \label{fig:init_dist}
\end{figure}

In Figure \ref{fig:init_dist}, we compare the initial semi-major axes calculated for the 40~pc sub-sample to those for our full sample of binaries. 
The initial semi-major axes of the 40~pc sub-sample tend to skew toward closer separations than the full binary sample. 
Within the 40~pc sub-sample, $\approx64-88\%$ of binary pairs have calculated initial separations within 1000~au, compared to $\approx66-76\%$ for the full WD/MS sample. 
This likely reflects the difficulty of resolving closer binaries with faint WDs, where the MS star may overshadow the WD, resulting in an under-sampling of close binaries in the full \textit{Gaia} sample. 

For the full sample, a total of $\approx24\%$ of binaries in each of the three different atmosphere model cases should be aligned according to our initial semi-major axis and companion mass criteria. The results for the three atmosphere models differ by at most $0.2\%$. 

For the 40~pc sub-sample, we expect $25-28\%$ of systems to be aligned, depending on the WD mass used (H, He, mixed, and corrected). This is slightly higher than the $24\%$ alignment fraction that we calculate for the full binary sample. Given the potential to miss distant but closely-separated WD/MS binaries in magnitude-limited stellar surveys, the alignment fraction of $25-28\%$ calculated for the 40~pc sub-sample may be more indicative of the overall fraction of WD/MS binaries across the galaxy in which ZLK oscillations may be prohibited by binary-driven dissipative precession. 

\subsection{ZLK timescale limits}\label{section:zlk_results}

\begin{figure}
    \centering
    \includegraphics[width=\linewidth]{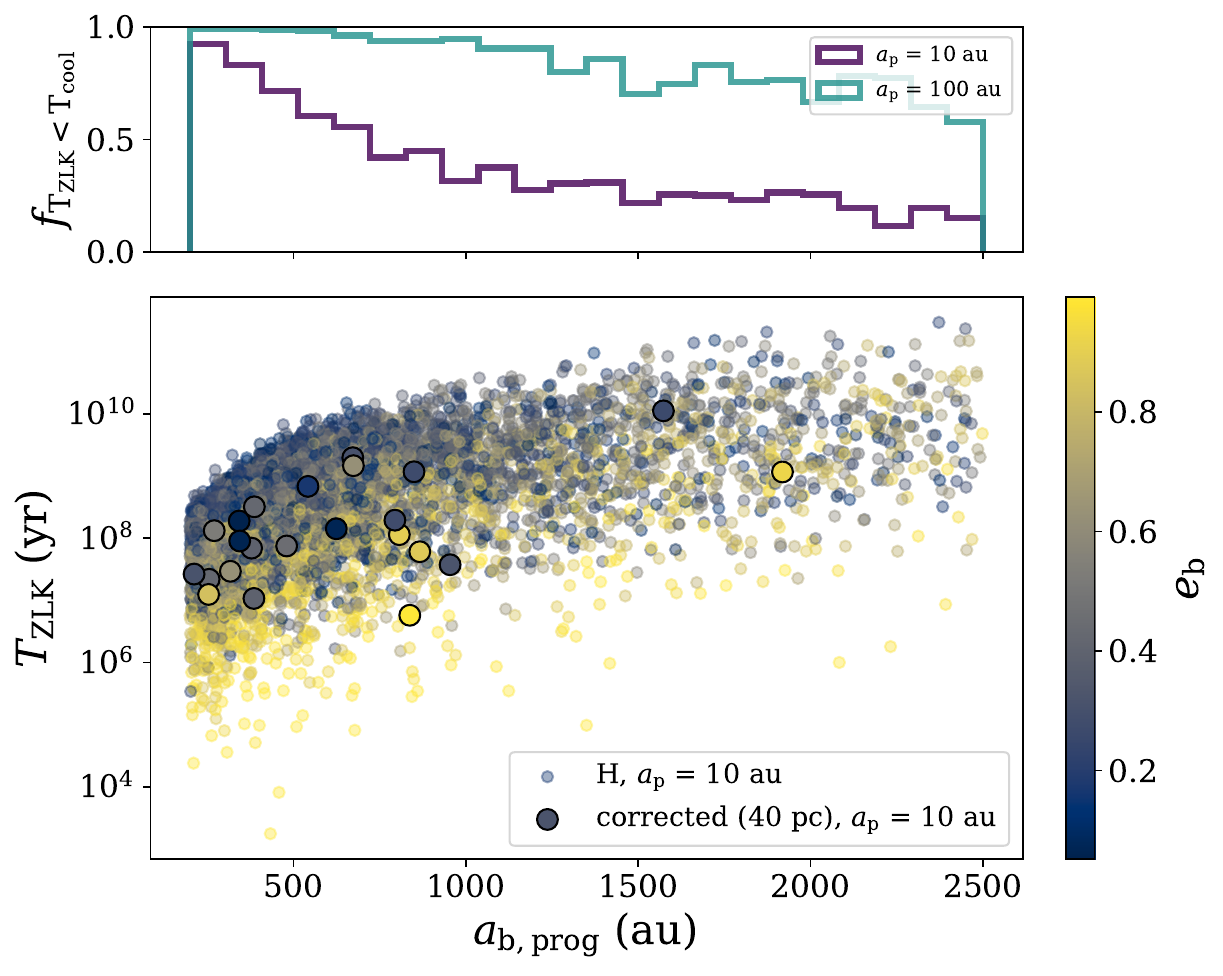}
    \caption{Distribution of calculated ZLK timescales (Eq.~\ref{eq:kozaitime}) for the WD/MS binaries in the full sample (small points) and 40~pc sub-sample (large points), colored by the systems' randomly selected eccentricity values. For clarity we show only the values calculated using the H, with corrected WD masses for the full sample and 40~pc sub-sample, respectively. The top panel shows the fraction of systems with $T_{\rm ZLK}<T_{\rm cool}$ in each bin of initial semi-major axis, for a polluting body at 10 au (purple) and 100 au (teal).}
    \label{fig:Tk_dist}
\end{figure}

Figure \ref{fig:Tk_dist} shows the distribution of ZLK timescales for the $a_{\rm p}=10$~au case, including the full binary sample and the 40~pc sub-sample. The distribution is colored by the random eccentricity assumed for each system. In the top panel of Figure \ref{fig:Tk_dist}, we show the fractions of binaries in each bin of $a_{\rm b, prog}$ whose ZLK timescales are shorter than the cooling age of the WD, for both the $a_{\rm p}=10$~au and $a_{\rm p}=100$~au cases.

ZLK timescales for the 40~pc sub-sample tend to skew shorter than those for the full sample, reflective of the higher fraction of binaries at close separations in the 40~pc sub-sample due to observational biases. 
The cooling ages of the 40~pc sample also tend to be higher than the those of the full sample.  Overall, $\approx60\%$ of the full sample and $\approx90\%$ of the sub-sample binaries have $T_{\rm ZLK}<T_{\rm cool}$ for the closer polluting body, while for the more distant polluter, $\approx94\%$ of the full sample and all of the 40~pc sub-sample could be polluted via ZLK-driven accretion processes in the past if alignment conditions are conducive to oscillations. 

The characteristic timescale of ZLK oscillations is therefore a strong restriction on the likelihood to accrete pollution if pollutable bodies generally originate in the inner solar system, but is typically not a limiting factor for more distant bodies. 

% \begin{figure}
%     \centering
%     \includegraphics[width=\linewidth]{Tk_dist.pdf}
%     \caption{Distribution of calculated ZLK timescales (Equation \ref{eq:kozaitime}) for the WD/MS binaries in the full sample (solid curves) and 40~pc sub-sample (dashed curve, using the ``corrected" WD masses). As the 40~pc sub-sample is smaller than the full sample, we multiply the counts for the 40~pc curve by 100 for visual clarity. The vertical line at $10^9$~yr denotes our imposed limit on systems that could be polluted by ZLK oscillations within typical WD cooling ages.}
%     \label{fig:Tk_dist}
% \end{figure}

\subsection{Modeled pollution frequency}\label{section:modeled_pollution}

We now combine the constraints of the ZLK timescale (Eq.~\ref{eq:kozaitime}) and alignment probabilities (Eq.~\ref{eq:prob_func}) to assess the fraction of WD/MS binary pairs that could have been polluted via ZLK-driven accretion according to our criteria outlined in Section \ref{section:pollution_criteria}. 
There are broadly two regimes of behavior: for binaries with $a_{\rm b, prog}\lesssim800$~au, alignment via dissipative precession is a limiting factor for ZLK pollution, while for $a_{\rm b, prog}\gtrsim800$~au pollution is only limited by the ZLK timescales relative to the WD cooling ages. 
Our model for dissipative precession predicts that the fraction of systems that can undergo ZLK oscillations (based on their alignment) should increase with initial semi-major axis until about 800~au, beyond which dissipative precession ceases to perturb the geometry of the system and final mutual inclinations are uniformly distributed. However, the ZLK timescale steeply increases with semi-major axis, thus reducing the number of systems that can be polluted. 

\begin{figure*}
    \centering
    \includegraphics[width=1\linewidth]{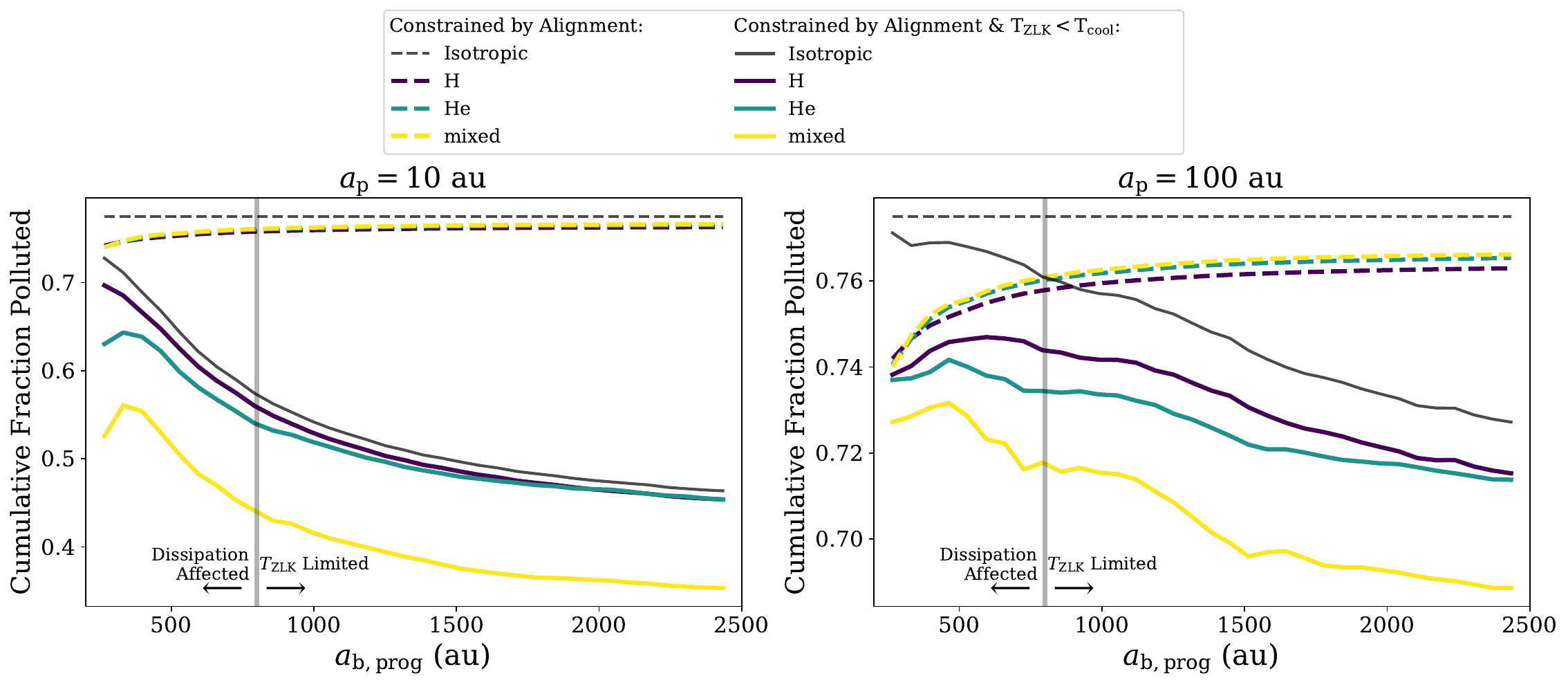}
    \caption{The trend in the cumulative fraction of binaries our model predicts could experience ZLK-driven pollution, as a function of initial semi-major axis, for the scenarios of pollution from an inner solar system body at 10~au during the WD phase (left) versus an outer solar system body at 100~au (right). 
    The dashed lines show the constraints on pollution fractions using only the alignment criteria, while the solid lines show the combined effects of alignment and ZLK timscale constraints. Curves are colored according to the atmospheric model assumed for the WD mass. For comparison, we show the isotropic curves in grey, demonstrating the expected pollution trend if dissipative precession is not active. 
    Dissipative precession can increase alignment fractions within about 800~au, beyond which ZLK timescales limit pollution. ZLK timescales are much shorter for more distant polluting bodies, allowing for higher potential rates of binary ZLK-induced pollution. }
    \label{fig:cumulative_models}
\end{figure*}

In Figure \ref{fig:cumulative_models}, we combine the alignment and ZLK timescale constraints to calculate the cumulative fraction of WD/MS systems in our sample that could have experienced ZLK-induced accretion after the formation of the WD, for both close ($a_p=10$ au) and distant ($a_p=100$ au) polluting bodies. 
While the trends in the pollution fractions are similar for the three atmospheric models explored, the `mixed' masses result in relatively less pollution, as their younger cooling ages set stricter limits on $T_{\rm ZLK}$ (see Section \ref{section:kozai_time}). 

The resulting trend predicts maximum binary-induced pollution frequencies for systems with initial binary semi-major axes within $\approx$500-800~au, depending on the distance of the polluting body. 
Dissipative precession reduces the occurrence of polluted systems through preferential alignment at the closest separations. 
Overall, if WDs mostly accrete inner solar system bodies, we expect up that no more than $50\%$ of the WDs in WD/MS binaries could have accreted ZLK-driven pollution in the past.
If accreted bodies are mostly distant solar system objects, we expect this upper limit to increase to about $70\%$. 

\section{Discussion}\label{section:discussion}

\subsection{Comparison to observed polluted white dwarfs}

\begin{figure}
    \centering
    \includegraphics[width=1\linewidth]{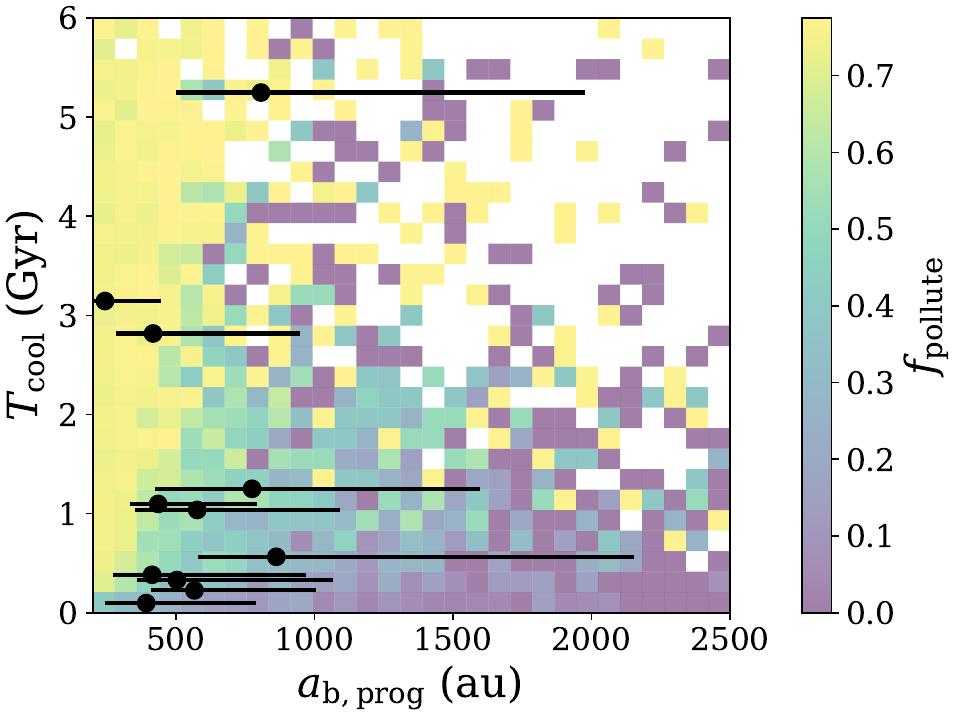}
    \caption{Comparison between the predicted fraction of WDs in our sample that have experienced ZLK-driven pollution, according to our model, and the binary WDs that have observed Ca pollution (black points). In this comparison, we assume the most restrictive case of a H atmosphere and a 10 au polluting body. The uncertainties in $a_{\rm b, prog}$ are primarily driven by the uncertainty in present-day orbital elements for each binary pair.  }
    \label{fig:data_comp}
\end{figure}

We now compare the predictions of our model to the small sub-sample of polluted WDs in binaries. We focus on the H atmosphere case for a 10~au polluting body as the most restrictive scenario. 
In Figure \ref{fig:data_comp}, we show the fraction of polluted WDs predicted by our model as a function of the initial semi-major axis of the binary and cooling age of the WD. 
The combined pollution criteria of inclined orbits ($\theta_{\rm db}(t = t_\mathrm{life})<39.2^\circ$) and sufficiently short ZLK timescales ($T_{\rm ZLK}<T_{\rm cool}$) result in the highest fractions of ZLK-polluted WDs for older WDs originating in more closely separated binaries.  

For comparison, we overplot the calculated initial semi-major axes and ages for the 11 known polluted WDs in binaries (Section \ref{methods:sample}). For each WD, we calculate the initial semi-major axis as in Section \ref{section:methods}, propagating uncertainties in the WD mass and companion mass. We use bootstrapping to account for uncertainty in present-day projected orbital elements, assuming that eccentricity and all relevant angles follow random uniform distributions.  

The resulting median $a_{\rm b, prog}$ for each WD is within 1000~au, with typical standard deviations of about 500~au.  
3/11 WDs fall in the parameter space of $T_{\rm cool}/a_{\rm b, prog}$ most favorable for high pollution frequencies ($>70\%$), where we expect the majority of WDs in binaries to have experienced ZLK-driven pollution in the past. 
The rest of the WDs are in the regions where we expect about $40-60\%$ of binary WDs to have undergone ZLK pollution. 

In the scenario of a polluting body with a 100~au orbit, pollution is only weakly timescale-limited. In this case, we expect pollution fractions of up to $80\%$ even for the more widely separated binaries, placing the observed polluted WDs well within this region. 

\subsection{Eccentricity distribution of stellar binary systems}\label{section:binary_eccentricity}

In this work, we calculate the semi-major axes of WD/MS binary orbits from their projected separations under the assumption that their eccentricities are uniformly distributed. We then extrapolate this present-day semi-major axis to the semi-major axis of the system prior to WD formation, assuming no change in the binary system's orbital eccentricity during the evolution and mass loss of the WD component. 

While eccentricities of close binaries (including all stellar types, and separations between a few and hundreds of au) have been measured to follow a uniform distribution, the eccentricity distribution gradually steepens to become a thermal, and eventually superthermal, distribution beyond $\sim10^3$~au \citep{Tokovinin2020,Hwang2022}. \cite{Hwang2025} find that present-day WD/MS binaries, in particular, have systematically lower eccentricities than those of the overall population of binaries. This may be because the WD/MS binaries form at close separations, and thus relatively lower eccentricities, and maintain their eccentricity after they widen during formation of the WD.  

% As the trend in WD/MS eccentricity distributions remain uncertain across the full range of separations considered in this work, and it is currently unclear if all WD/MS binaries would maintain their initial eccentricities throughout formation and mass loss of the WD, we do not implement a separation-dependent eccentricity distribution in this work. 
 
If wide-separation WD/MS binaries have higher eccentricities than assumed in this work, we would expect that their calculated present-day semi-major axes (Eq.~\ref{equation:aMS}) would be larger than we calculate with a uniform eccentricity distribution, while their ZLK timescales would be shorter (Eq.~\ref{eq:kozaitime}). This would increase the total number of systems that could experience ZLK-induced pollution. 

\subsection{Considering WD/WD binary systems}
In this work we focus on WDs in a binary system with a MS companion. However, similar reasoning could be applied to double WD binaries, in which one WD would drive pollution onto its companion. 
Observed WD/WD binaries are much less prevalent than WD/MS systems; cross-referencing the \cite{El-Badry2021} and \cite{GentileFusillo2021} catalogs, we find  854 WD/WD pairs. 
Using the same methods as outlined in Section \ref{methods:initialconditiona}, we calculate the present-day semi-major axes, progenitor masses for both WD components, and initial semi-major axes for each of the double WD systems. 
To account for the evolution of both WDs in each binary, we solve for the initial semi-major axis in two steps, assuming that the more massive component evolved into a WD before the less massive component. 

The WD/WD binaries tend to be more closely separated than the WD/MS sample explored in this work.
The calculated initial semi-major axes of the WD/WD systems have a median value of about 50~au, and over $80\%$ of the binary pairs have initial semi-major axes within 200~au. 
The systems' small initial semi-major axes suggest that these systems are unlikely to undergo ZLK-driven pollution, as the bulk of the sample falls into the regime of potentially suppressed planet formation and/or strong preferential alignment via dissipative precession. 

More widely-separated WD/WD binaries could still be candidates for ZLK-induced pollution, though these systems are more rare. 
The dearth of widely-separated WD/WD binaries, compared to WD/MS and MS/MS systems, has been pointed out in the broader \textit{Gaia} binary census as well as in the 40~pc WD survey \citep{El-Badry2018, McCleery2020, O'Brien2024}. 
The rarity of such systems may result from wider binaries being disrupted during stellar evolution, or from dynamical interactions with other stars \citep[e.g.,][]{Toonen2017, El-Badry2018, Torres2022}.
Overall, ZLK-induced WD pollution appears to be much less likely for WD/WD binary systems than for their WD/MS counterparts.  

\subsection{Implications for pollution rates at present day}

Our model predicts the fraction of binary WD/MS systems that could have undergone ZLK-induced pollution in the past. 
Translating this value to the fraction of binary WD/MS systems that should display WD polluted at present day requires several assumptions regarding the total number and rate of bodies in a given WD system that can be perturbed by ZLK oscillations, as well as the timescale of tidal disruption and the typical duration of an accretion event before the material sinks out of the WD atmosphere.

\cite{Stephan2017} propose eccentric ZLK oscillations as a means of delivering individual massive, volatile-rich polluters to WDs, as opposed to the more continuous accretion expected from planetary perturbations of debris \citep{Mustill2018, O'Connor2022, O'Connor2023}.
To test a simple analogue of this model, we therefore assume that each WD in our sample accretes a single body at the calculated time $T_{\rm ZLK}$ whose material remains in the atmosphere for $\sim1$~Myr (a typical settling time for He-dominated WDs, \citealt{Bedard2020}). Requiring the cooling age to be within $\sim1$~Myr of $T_{\rm ZLK}$, we predict that only 3 WDs out of the full sample of 4400 should host observable material at their present cooling ages. 

A low occurrence of observed present-day ZLK-driven pollution would be consistent with previous work, including \cite{Noor2024}, who find that pollution rates among WDs are independent of the presence of wide stellar binary companions, and \cite{Wilson2019}, who find that the presence of pollution and accretion disks, identified via infrared excesses, does not differ between single and binary samples. 
It is therefore possible that, while the conditions required for ZLK-induced pollution are common among binary WD/MS systems, the relative infrequency of such individual pollution events compared to continuous planetary-driven pollution restricts their impact on overall, observed pollution rates.

% \cite{Jenkins2024} find that the metallicities of WD companions, indicative of the WD progenitor metallicity and thus correlating with the occurrence rates of giant planets around WD progenitors, are not correlated with pollution rates, suggesting that giant planets do not drive accretion of pollution. Dynamical simulations also support the theory of lower-mass planet perturbers, and find that sub-Neptune mass planets beyond 10~au are capable of providing long-term pollution to WDs \citep{Mustill2018, O'Connor2022, O'Connor2023}. It is possible that such planets are more likely to be in high-multiplicity systems to induce instabilities more quickly \citep{Maldonado2021}. 

% On the extremely low-mass end, bodies as small as the Moon may also be capable of inducing pollution \citep{Veras2023}. As a final alternative, it is possible to induce pollution without any planet or stellar binary dynamics,  through mechanisms such as radiation-driven orbital decay \citep{Veras2022} or dynamical kicks from anisotropic mass loss during WD formation \citep{Akiba2024}. 

\section{Conclusions}\label{section:conclusion}

In this work, we test the prevalence of ZLK-driven WD pollution, considering the impact of binary system alignment due to dissipative precession of protoplanetary disks. Our sample includes polluted and unpolluted WDs each with a MS stellar companion, where each binary system was identified through a cross-match of catalogues derived from \textit{Gaia} astrometry \citep{El-Badry2021, GentileFusillo2021}. We consider the full set of observed WD-MS binaries in this range, as well as the subset included in the complete, volume-limited 40~pc sample of WDs described in \cite{O'Brien2024}. 

We consider two criteria to constrain the frequency of ZLK-driven pollution: first, systems must have mutual inclinations $>39.2^\circ$ to allow ZLK oscillations to occur, and second, the characteristic timescale for ZLK oscillations must be shorter than the cooling age of the WD.

We first model the degree of alignment expected for WD/MS binaries in the \textit{Gaia} survey, incorporating the impact of dissipative precession in shaping the early evolution of the system geometries. 
Dissipative precession is most effective at aligning binaries with small initial semi-major axes, and it can thus prevent ZLK oscillations in up to $30\%$ of close-separation binaries (within 500~au), depending on the properties and configuration of the natal protoplanetary disk and companion star. 
The effect of dissipative precession weakens beyond 800~au, resulting in about $24\%$ of observed \textit{Gaia} WD/MS pairs being sufficiently aligned (mutual inclination $<39.2^\circ$) to prevent ZLK oscillations.

We further limit the number of binaries in our sample that could have experienced ZLK-driven pollution by requiring that the ZLK timescales must be shorter than the WD cooling age. 
We calculate separate ZLK timescales for the cases of pollution from the inner system (10~au perturber orbit around the WD) and outer system (100~au perturber orbit).
For the close and distant cases, our models predict upper limits of $50\%$ and $72\%$, respectively, for the subset of WD systems that could have undergone ZLK-driven pollution. 

Our model suggests that the conditions necessary for ZLK-driven pollution cannot be ruled out for most WD/MS binaries with initial semi-major axes of 200-2500~au. 
We find that dissipative precession limits ZLK oscillations for the more closely separated pairs (within 800~au), while characteristic ZLK timescales limit pollution for more widely separated binaries. 

Despite the prevalence of favorable conditions for ZLK pollution, observed rates of pollution for WDs in binaries remain very low ($0.2\%$ in our sample and $2-10\%$ in the 40~pc sub-sample, \citealt{O'Brien2024}) compared to the $30-50\%$ of WDs that are polluted overall \citep{Koester2014}. 
This may be explained by the nature of ZLK-driven pollution itself, as ZLK oscillations could perturb fewer, more massive polluting bodies compared to planetary drivers, resulting in a low observed  pollution rate in such systems even under the influence of ZLK dynamics.

% Comparing to polluted WDs in WD/MS pairs in \textit{Gaia}, we find that the cumulative fraction of polluted binary WDs reaches a stable limit for initial binary separations $\gtrsim500$~au. While our modeled trend at wider separations ($\gtrsim1000$~au) flattens as ZLK timescales increase, the observed pollution rates diverge from our model at smaller separations, where rates of ZLK-induced pollution should trend upwards. Thus, while ZLK oscillations may be active in some cases, binary-independent dynamical scattering appears to dominate pollution at least for more closely separated binaries. 

\begin{acknowledgments}
We thank the Rice Research Group for helpful conversations over the course of this work. We also thank the referee for constructive feedback. This work was supported by Heising-Simons Foundation Grants \#2021-2802 and \#2023-4478, as well as National Geographic Grant EC-115062R-24.

\software{NumPy \citep{Harris2020}, SciPy \citep{Virtanen2020}, Matplotlib \citep{Hunter2007}, CMasher \citep{vanderVelden2020}}

\end{acknowledgments}

\appendix

\section{Modeling dissipative precession in the progenitor binary}
\label{section:appenidx_dissipative_precession}

We consider a WD progenitor with spin angular momentum vector $\bm{S} = S\hat{\bm{s}}$ that hosts a disk with angular momentum vector $\bm{L}_\mathrm{d} = L_\mathrm{d}\hat{\bm{l}}_\mathrm{d}$, where
\begin{align}
    S &= k_*  \bar{\Omega}_*M_\mathrm{prog}\sqrt{GM_\mathrm{prog}R_\mathrm{prog}},\\
    L_\mathrm{d} &= \frac{2 - p}{5/2 - p }M_\mathrm{d}\sqrt{GM_\mathrm{prog} r_\mathrm{out}}.
\end{align}
For a fixed binary orbital angular momentum $\bm{L}_\mathrm{b}/\hat{\bm{l}}_\mathrm{b} = L_\mathrm{b} \gg S,L_\mathrm{d}$, the unit vectors of the WD progenitor's spin and disk co-evolve via \citep[][]{Zanazzi2018}
\begin{align}
\label{eq:ld_evol}
\begin{split}
    \frac{\dif \hat{\bm{l}}_\mathrm{d}}{\dif t} = &- \tilde{\omega}_\mathrm{ds}( \hat{\bm{l}}_\mathrm{d} \cdot  \hat{\bm{s}})  \hat{\bm{s}} \times  \hat{\bm{l}}_\mathrm{d} - \tilde{\omega}_\mathrm{db}( \hat{\bm{l}}_\mathrm{d} \cdot  \hat{\bm{l}}_\mathrm{b})  \hat{\bm{l}}_\mathrm{b} \times  \hat{\bm{l}}_\mathrm{d} + \gamma_\mathrm{b}(\hat{\bm{l}}_\mathrm{d}\cdot \hat{\bm{l}}_\mathrm{b})^3 \hat{\bm{l}}_\mathrm{d} \times (\hat{\bm{l}}_\mathrm{b} \times \hat{\bm{l}}_\mathrm{d}) + \gamma_\mathrm{s}(\hat{\bm{l}}_\mathrm{d}\cdot \hat{\bm{s}})^3 \hat{\bm{l}}_\mathrm{d} \times (\hat{\bm{s}} \times \hat{\bm{l}}_\mathrm{d}),
\end{split}\\
\label{eq:2_evol}
 \frac{\dif \hat{\bm{s}}}{\dif t} = & -\tilde{\omega}_\mathrm{sd} (\hat{\bm{s}}\cdot \hat{\bm{l}}_\mathrm{d}) \hat{\bm{l}}_\mathrm{d} \times \hat{\bm{s}} - \frac{L_\mathrm{d}}{S}\gamma_\mathrm{s}(\hat{\bm{l}}_\mathrm{d}\cdot \hat{\bm{s}})^2 \hat{\bm{s}} \times (\hat{\bm{s}} \times \hat{\bm{l}}_\mathrm{d}),
\end{align}
where the precession rates are given by
\begin{align}
    \tilde{\omega}_\mathrm{ds} &= \frac{3(5/2 - p)k_\mathrm{q}}{2(1+p)}\frac{R_\mathrm{prog}^2 \bar{\Omega}_*^2}{r_\mathrm{out}^{1-p}r_\mathrm{in}^{1+p}} \sqrt{\frac{GM_\mathrm{prog}}{r_\mathrm{out}^3}},\\ 
    \tilde{\omega}_\mathrm{db} &= \frac{3(5/2 - p)}{4(4-p)}\left(\frac{M_\mathrm{C}}{M_\mathrm{prog}}\right)\left(\frac{r_\mathrm{out}}{a_\mathrm{b,eff}}\right)^3 \sqrt{\frac{GM_\mathrm{prog}}{r^3_\mathrm{out}}}, \label{eq:omdb} \\ 
    \tilde{\omega}_\mathrm{sd} &= \frac{3(2 - p)k_\mathrm{q}}{2(1+p)k_*}\left(\frac{M_\mathrm{d}}{M_\mathrm{prog}}\right) \frac{\bar{\Omega}_*\sqrt{GM_\mathrm{prog}R_\mathrm{prog}^3}}{r_\mathrm{out}^{2-p}r_\mathrm{in}^{1+p}}.
\end{align}
The damping rates scale as
\begin{align}
\begin{split}
\label{eq:damping_rate_binary_warps}
    \gamma_\mathrm{b}  =  & 1.26 \cdot 10^{-9} \left(\frac{\alpha}{10^{-2}}\right)\left(\frac{h_\mathrm{out}}{0.1}\right)^{-2} \left(\frac{M_\mathrm{C}}{1 M_\odot}\right)^2  \cdot\left(\frac{r_\mathrm{out}}{50 \mathrm{au}}\right)^{\frac{9}{2}}  \left(\frac{a_\mathrm{b,eff}}{300 \mathrm{au}}\right)^{-6} \left(\frac{M_\mathrm{prog}}{1 M_\odot}\right)^{-\frac{3}{2}} \frac{2\pi}{\mathrm{yr}},
\end{split}\\
\begin{split}
\label{eq:damping_rate_oblate_star_warps}
    \gamma_\mathrm{s}  = & 2.04 \cdot 10^{-10} \left(\frac{\alpha}{10^{-2}}\right)\left(\frac{h_\mathrm{in}}{0.1}\right)^{-2} \left(1358\frac{r_\mathrm{out}}{r_\mathrm{in}}\right)^{p-1}  \left(\frac{k_\mathrm{q}}{0.1}\right)^2 \left(\frac{\bar{\Omega}_*}{0.1}\right)^{4} \left(\frac{M_\mathrm{prog}}{1 M_\odot}\right)^{\frac{1}{2}} \left(\frac{R_\mathrm{prog}}{2 R_\odot}\right)^{4}   \left(\frac{r_\mathrm{in}}{8 R_\odot}\right)^{-4} \left(\frac{r_\mathrm{out}}{50 \mathrm{au}}\right)^{-\frac{3}{2}} \frac{2\pi}{\mathrm{yr}}.
\end{split}
\end{align}
In the above expressions, $p$ is the power law index of the disk surface density profile ($\Sigma(r) \propto r^{-p}$), $\bar{\Omega}_*$ is the dimensionless rotation rate of the progenitor, and $k_*$ and $k_\mathrm{q}$ are the dimensionless progenitor spin and progenitor quadropole moment normalizations, respectively. For all calculations, we fix $p = 1$, $\bar{\Omega}_* = 0.1$, $k_* = 0.2$, and $k_\mathrm{q} = 0.1$. $h_\mathrm{in}$ and $h_\mathrm{out}$ are the disk aspect ratio at the inner and outer truncation radius, respectively. For these, and for viscosity parameter $\alpha$ and progenitor radius $R_\mathrm{prog}$, we employ the scaling relations discussed in Sect.~\ref{section:simulation_methods}.

In this paper, we are interested in the evolution of the mutual inclination between disk and binary orbit $\theta_\mathrm{db} = \arccos(\hat{\bm{l}}_\mathrm{d} \cdot \hat{\bm{l}}_\mathrm{b})$. Specifically, we evolve the dynamical system in Eq.~\eqref{eq:ld_evol} and \eqref{eq:2_evol} forward in time as in \citet{Gerbig2024} and evaluate $\theta_\mathrm{db}(t=t_\mathrm{life})$ to obtain the predicted `final' inclination for some given initial inclination $\theta_\mathrm{db}(t=0)$. Motivated by expectations from binary star formation via turbulent fragmentation, we expect the disk angular momentum vector to initially match the progenitor spin vector, and for the initial mutual inclination between the disk and the binary orbit to follow a random uniform distribution such that $P[\theta_\mathrm{db}(t=0)] \propto \sin |\theta_\mathrm{db}(t=0)|$ \citep[e.g.][]{bate2010chaotic}. We therefore sample uniformly from $P$ to evolve an ensemble of systems with identical physical parameters. We ultimately obtain an expectation for the distribution of $P[\theta_\mathrm{db}(t=t_\mathrm{life})]$ that is agnostic to the initial orientation of the system.

\bibliography{bibli}{}
\bibliographystyle{aasjournal}

\end{CJK}
\end{document}